\documentclass[reprint,amsmath,amssymb,aps,prb,showpacs,floatfix,superscriptaddress]{revtex4-1}
\usepackage{graphicx}
\usepackage{bm}
\usepackage{multirow}
\usepackage{verbatim}
\usepackage{longtable}
\usepackage{color}
\usepackage{nccmath}

\usepackage[normalem]{ulem}
\renewcommand{\figurename}{Fig.}

\newcommand{\sectionname}{Section}

\newcommand{\e}{e}
\renewcommand{\i}{i}
\newcommand{\mc}{\mathcal}
\newcommand{\mr}{\mathrm}

\newcommand{\ve}{\varepsilon}

\newcommand{\pd}{\partial}
\newcommand{\dif}{\mathrm{d}}

\newcommand\ph[1]{\phantom{#1}}
\newcommand\abs[1]{\lvert#1\rvert}
\newcommand\absB[1]{\left\lvert#1\right\rvert}

\newcommand\bra[1]{\langle#1\rvert}
\newcommand\ket[1]{\lvert#1\rangle}

\newcommand{\Imag}{\operatorname{Im}}

\newcommand{\Tr}{\operatorname{Tr}}

\newcommand{\up}{\uparrow}
\newcommand{\down}{\downarrow}

\newcommand{\cd}{c^{\dag}}
\newcommand{\can}{c^{\phantom{\dag}}}

\newcommand{\dd}{d^{\dag}}
\newcommand{\dan}{d^{\phantom{\dag}}}

\newcommand{\shs}[1]{}
\newcommand{\kap}{\kappa}

\begin{document}

%-----------------------------------------------------------%
\title{Thermopower signatures and spectroscopy of the canyon of conductance suppression}

\author{G. Kir{\v{s}}anskas }
\affiliation{Mathematical Physics and NanoLund, University of Lund, Box 118, 22100 Lund, Sweden}

\author{S. Hammarberg}
\affiliation{Mathematical Physics and NanoLund, University of Lund, Box 118, 22100 Lund, Sweden}

\author{O. Karlstr{\"o}m}
\affiliation{Mathematical Physics and NanoLund, University of Lund, Box 118, 22100 Lund, Sweden}

\author{A. Wacker }
\affiliation{Mathematical Physics and NanoLund, University of Lund, Box 118, 22100 Lund, Sweden}

\date{\today}
%-----------------------------------------------------------%

%-----------------------------------------------------------%
\begin{abstract}
Interference effects in quantum dots between different transport channels can lead to a strong suppression of conductance, which cuts like a canyon through the common conductance plot [Phys.\ Rev.\ Lett.~{\bf 104}, 186804 (2010)]. In the present work, we consider the thermoelectric transport properties of the canyon of conductance suppression using the second-order von Neumann approach. We observe a characteristic signal for the zeros of the thermopower. This demonstrates that thermoelectric measurements are an interesting complimentary tool to study complex phenomena for transport through confined systems.
\end{abstract}
%-----------------------------------------------------------%

\maketitle

\section{Introduction}

Thermoelectric effects in nanoscale structures have been frequently studied with the aim to improve the efficiency of devices.~\cite{DresselhausAdvMater2007,SnyderNatMater2008} In addition, it has been recently demonstrated that thermopower measurements can serve as an interesting tool to characterize complex scenarios due to coherences and interactions in nanoscale systems.~\cite{ScheibnerPRB2007,ScheibnerNJP2008,SwirkowiczPRB2009,CostiPRB2010,AndergassenPRB2011,RejecPRB2012,SvenssonNJP2013,LeijnseNJP2014,MatthewsPRB2014} Here the intrinsic advantage of thermoelectric measurements is that they probe asymmetries around the Fermi level and therefore can easily provide information about excited states.~\cite{KristinsdottirAPL2014}

Close to a degeneracy of energy levels in a quantum dot, interference and correlation effects can play an important role for transport and lead to pronounced quantum mechanical phenomena. For a spin-degenerate quantum dot (QD) the conductance experiences an enhancement for low temperatures due to the Kondo effect.~\cite{GoldhaberNature1998, CronenwettScience1998}
The thermopower of the single and multiple quantum dots in the presence of the Kondo effect was examined both theoretically~\cite{
BoeseEPL2001,
DongJPhysCondensMat2002,
KimPRL2002,
KrawiecPRB2006,
SakanoJPSJ2007,
ZhangJPhysCondensMat2007,
FrancoJAP2008,
YoshidaPHB2009,
CostiPRB2010,
NguyenPRB2010,
AzemaPRB2012,
Roura-BasPRB2012,
KarwackiJPhysCondensMat2013,
WeymannPRB2013,
YePRB2014}
and experimentally.~\cite{ScheibnerPRL2005}
On the other hand, in the case of two degenerate levels with equal spin, at degeneracy, conductance can be suppressed due to electron correlation and interference effects.~\cite{ShahbazyanPRB1994,KubalaPRB2002,GuevaraPRB2003,MedenPRL2006,KashcheyevsPRB2007,TokuraNJP2007} Such a spin-polarized two-level model was widely used to interpret the phenomena of phase lapses of transmission in Aharonov-Bohm interferometer containing a QD.~\cite{YacobyPRL1995,OregPRB1997,SchusterNature1997,Avinun-KalishNature2005,KarraschNJP2007,KarraschPRL2007a,KashcheyevsPRB2007} Additionally, due to their small size the quantum dots also can exhibit Coulomb blockade effect.~\cite{AverinJLowTempPhys1986,GrabertBook1992} The thermopower of the usual Coulomb blockade sequential tunneling peaks and the cotunneling signal were addressed in Refs.~\onlinecite{BeenakkerPRB1992,CobdenPHA1993,DzurakSolidStateCommun1993,StaringEPL1993,
DzurakPRB1997,MollerPRL1998,AndreevPRL2001,MatveevPRB2002,TurekPRB2002}.

In Ref.~\onlinecite{NilssonPRL2010} a system of two spin-polarized degenerate levels was realized in an InSb nanowire QD, where different $g$-factors allow to control level crossings for the same spin by a magnetic field. The experiment showed, in good agreement with supporting calculations, that the suppression cuts as a canyon through the standard conductance plot for different parities of the level couplings. A more detailed analysis of the conductance spectrum can be found in Ref.~\onlinecite{KarlstromPRB2011a}. A related two-level system was optimized for achieving high thermoelectric performance in Ref.~\onlinecite{KarlstromPRB2011b}. In this work we further elaborate on thermoelectric properties and focus on the fingerprint of conductance suppression in the thermopower signal.

\begin{figure}
\begin{center}
\includegraphics[width=0.9\columnwidth]{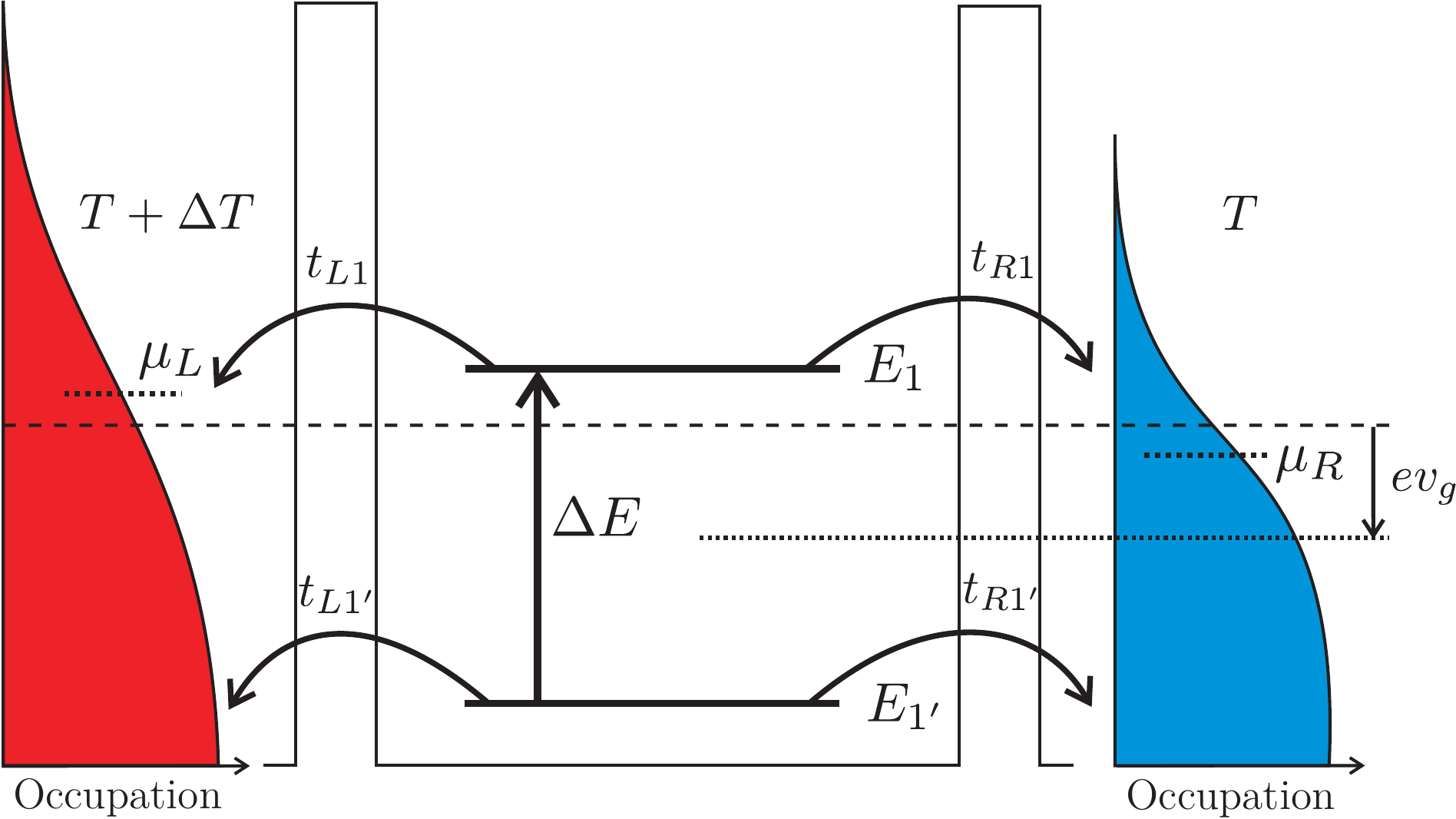}
\caption{\label{fig1} (Color online) Schematic of the system: A quantum dot with single
  particle levels $E_1$ and $E_{1'}$ coupled to the leads via tunneling
  barriers. The leads have a temperature difference $\Delta{T}$
  which can give rise to a current flowing through the dot. }
\end{center}
\end{figure}

The paper is organized as follows. In \sectionname~\ref{sec:model} the model for the spin-polarized two-level quantum dot is introduced. Results for conductance and thermopower are presented in \sectionname~\ref{sec:results}. Here we focus on the zeros of the thermopower, which are relatively easy to extract experimentally. We start in \sectionname~\ref{subsec:uzero} with a discussion of the non-interacting case, where results are obtained using transmission formalism. Here we show that up to five zeros in the thermopower can be found if the gate voltage is varied. Furthermore, we establish the role of temperature and level broadening for the existence of multiple zeros. The impact of finite QD charging energy is addressed in \sectionname~\ref{subsec:ufinite}, where calculations are performed by the second order von Neumann (2vN) approach. This reveals the full scenario for the canyon of conductance suppression. Concluding remarks are given in \sectionname~\ref{sec:concl}.

\section{\label{sec:model}Model}

The two-level quantum dot system under investigation is depicted in Fig.~\ref{fig1}. Its energy levels $E_1, \ E_{1'}$ are assumed to be individually tunable. In practice this is achieved by a gate voltage $v_g=(E_1+E_{1'}-\mu_{\mr{L}}-\mu_{\mr{R}})/2e$ (where $e<0$ is the electron charge, $\mu_{L/R}$ is the chemical potential in the left and right lead, respectively); and by controlling the detuning $\Delta{E}=E_1-E_{1'}$, which, for example, can be physically realised by a magnetic field if the two levels have different $g$-factors.\cite{NilssonPRL2010}

The total Hamiltonian for the system consisting of the QD coupled to two leads via tunneling barriers can be written as
\begin{equation}\label{hamFull}
H=H_{\mr{D}}+H_{\mr{LR}}+H_{\mr{T}}.
\end{equation}
The two-level spinless QD Hamiltonian $H_\mathrm{D}$, in a single particle basis, is
\begin{equation}\label{hamD}
H_{\mr{D}}= E_1^{\ph{\dag}}\dd_1\dan_1 + E_{1'}^{\ph{\dag}}\dd_{1'}\dan_{1'} + U\dd_1\dan_1\dd_{1'}\dan_{1'},
\end{equation}
where $U$ is the charging energy due to Coulomb repulsion between electrons in the dot. In our calculations and corresponding plots we shift the zero value of the gate voltage to the particle-hole symmetric point and use
\begin{equation}
V_{g}=\abs{e}v_{g}-U/2.
\end{equation}
The lead Hamiltonian $H_{\mr{LR}}$ is
\begin{equation}\label{hamLR}
H_{\mr{LR}}=\sum_{k,\ell}E_{k}^{\ph{\dag}}\cd_{k\ell}\can_{k\ell},
\end{equation}
where $k$ denotes the lead state and $\ell=\mr{L},\mr{R}$ denotes the lead. We assume, that the leads are thermalized resulting in an occupation function
\begin{equation}
f_\ell(k)=\frac{1}{e^{(E_{k}-\mu_\ell)/k_{\mathrm{B}}T_\ell}+1},
\end{equation}
with different temperatures $T_{\mr{L}/\mr{R}}$, for the left and right lead, respectively. The tunneling between the dot and the lead is governed by the Hamiltonian $H_{\mr{T}}$ that reads
\begin{equation}\label{hamT}
H_{\mr{T}}=\sum_{k,\ell}(t_{\ell1}^{\ph{\dag}}\dd_1 + t_{\ell1'}^{\ph{\dag}}\dd_{1'})\can_{k\ell} + \mathrm{h.c.},
\end{equation}
where, for simplicity, the tunneling amplitudes $t_{\ell i}$ are assumed to be the same for all values of $k$ and the coupling strengths are defined as $\Gamma_{\ell i}(E)=2\pi \abs{t_{\ell i}}^2 \sum_k \delta(E_k -E)\approx 2\pi\nu_{F} \abs{t_{\ell i}}^2$. Here we performed the $k$-sum using the flat density of states approximation, i.~e., $\sum_{k}\rightarrow \nu_{F}\int_{-D}^{D} \dif{E}$, with $\nu_{F}$ denoting the density of states at the Fermi level and $2D$ denoting the bandwidth of the leads. We also assume $2D$ to be the largest energy scale in the problem.

In a nonequilibrium situation a current $I$ through the dot can be generated both by an applied bias $V=(\mu_{\mr{L}}-\mu_{\mr{R}})/e$ and a temperature difference $\Delta T=T_{\mr{L}}-T_{\mr{R}}$ between the leads. Close to equilibrium this establishes a linear relation between $\Delta T$ and the bias $V$ for which this current vanishes. This defines the Seebeck coefficient (or thermopower) $S$ via
\begin{equation}
S=\left.-\frac{V}{\Delta T}\right|_{I=0}.
\end{equation}
This coefficient can be numerically obtained for a given transport model.
For the units of the thermopower we use $S_{0}=k_{\mathrm{B}}/\abs{e}\approx 86.1 \ \mu\mathrm{V/K}$ and for the units of conductance we use $G_{0}=e^2/h\approx 38.7 \ \mu\mathrm{S}$.
Lastly, in linear response to applied bias $V$ or temperature difference $\Delta{T}$ the Hamiltonian Eq.~\eqref{hamFull} can be mapped to generalized Anderson model, which was discussed in Ref.~\onlinecite{KashcheyevsPRB2007}. This mapping for model parameters considered in this paper is discussed in \appendixname~\ref{sec:mapAnderson}.

\subsection{Non-interacting case $U=0$}
For the non-interacting case, when there is no charging energy $U=0$, the thermopower $S$ and conductance $G$ can be calculated exactly from the electronic transmission $\mc{T}(E)$ as~\cite{SivanPRB1986,LundeJPhysCondensMatter2005,EsfarjaniPRB2006}
\begin{subequations}\label{GStransform}
\begin{align}
&\label{uezsigma}G=e^2L_{0},\\
&\label{uezS}S=-\frac{1}{\abs{e}T}\frac{L_1}{L_{0}},
\end{align}
\end{subequations}
with
\begin{equation}
L_{m}=\frac{1}{h}\int_{-\infty}^{+\infty}\dif{E}(E-\mu)^{m}\left(-\frac{\pd f(E,\mu,T)}{\pd E}\right)\mc{T}(E),
\end{equation}
where the limit of vanishing temperature difference $\Delta{T}\rightarrow0$ and vanishing bias $V\rightarrow0$ between the leads was taken.
The transmission is calculated using the Caroli formula~\cite{CaroliJPhysCSolidStatePhys1971}
\begin{equation}\label{P_VD_tflb}
\mc{T}(E)=\Tr[\Gamma_{\mr{L}}G^{R}(E)\Gamma_{\mr{R}}G^{A}(E)],
\end{equation}
where $G^{R/A}(E)$ is the retarded/advanced Green's function of the quantum dot electrons and $\Gamma_{\ell}$ is the coupling strength given by
\begin{equation}
\Gamma_{\ell}=\begin{pmatrix}
\Gamma_{\ell, 11} & \Gamma_{\ell, 11'} \\
\Gamma_{\ell, 1'1} & \Gamma_{\ell, 1'1'}
\end{pmatrix},\quad \Gamma_{\ell, ij}=2\pi\nu_{F}t_{\ell i}^{\ph{}}t_{\ell j}^{*}.
\end{equation}
The retarded/advanced Green's function can be obtained from
\begin{subequations}
\begin{align}
&G=(G_{0}^{-1}-\Sigma)^{-1},\\
%%%
&G_{0}^{-1}=\begin{pmatrix}
z-E_1 & 0 \\
0 & z-E_{1'}
\end{pmatrix},
\end{align}
by replacing $z=E+\i\eta$ for the retarded function $G^{R}$ and $z=E-\i\eta$ for the advanced function $G^{A}$, where $\eta$ is positive infinitesimal. Here $\Sigma$ is the self-energy given by
\begin{equation}
\Sigma=\Sigma_{\mr{L}}+\Sigma_{\mr{R}}, \quad
\Sigma_{\ell}^{R/A}=\mp\i\frac{\Gamma_{\ell}}{2},
\end{equation}
\end{subequations}
in the case of infinite bandwidth $D\rightarrow+\infty$.
If the following choice of tunneling amplitudes (corresponding to the canyon of conductance suppression) is made
\begin{equation}\label{amplitudes}
t_{\mr{L}1}=t, \ t_{\mr{R}1}=t, \ t_{\mr{L}1'}=-at, \ t_{\mr{R}1'}=at,
\end{equation}
we get the coupling strength matrices
\begin{equation}
\Gamma_{\mr{L}}=\Gamma\begin{pmatrix}
1 & -a \\
-a & a^2
\end{pmatrix},\quad
\Gamma_{\mr{R}}=\Gamma\begin{pmatrix}
1 & +a \\
+a & a^2
\end{pmatrix},
\end{equation}
with $\Gamma=2\pi\nu_{F}\abs{t}^2$, and the transmission function takes a simple and intuitive form as a sum of two Breit-Wigner resonances:~\cite{BreitPR1936,NakanishiJPSJ2007,Xu-MingChinesePhysLett2009,GongJPSJ2012}
\begin{equation}\label{transanl}
\mc{T}(E) = \Gamma^2 \absB{\frac{1}{E-E_1+i\Gamma} - \frac{a^2}{E-E_{1'}+ia^2\Gamma}}^2.
\end{equation}
Here the minus sign between both terms relates to the different parities of the tunnel couplings for both levels as defined in Eq.~(\ref{amplitudes}). For small $\Gamma$ both terms cancel at $E_{1'}=a^2E_1$ resulting in zero transmission. This provides a line of conductance suppression in the $(E_1,E_{1'})$ plane.

In order to see whether the canyon of conductance suppression gives qualitatively different results for the thermopower, we will make a comparison to the case when all the tunneling amplitudes have the same sign:
\begin{equation}\label{amplitudesS}
t_{\mr{L}1}=t, \ t_{\mr{R}1}=t, \ t_{\mr{L}1'}=+at, \ t_{\mr{R}1'}=at.
\end{equation}
This choice leads to such coupling strength matrices
\begin{equation}
\Gamma_{\mr{L}/\mr{R}}=\Gamma\begin{pmatrix}
1 & + a \\
+ a & a^2
\end{pmatrix}.
\end{equation}
The transmission function acquires a more complicated structure than a sum of two resonances:~\cite{ChoPRB2005,NakanishiJPSJ2007,Xu-MingChinesePhysLett2009,ZhengJLowTempPhys2012,GongJPSJ2012}
\begin{equation}\label{transanlS}
\mc{T}(E) = \Gamma^2 \absB{\frac{a^2(E-E_{1})+(E-E_{1'})}{(E-E_{1}+\i\Gamma)(E-E_{1'}+\i a^2\Gamma)+a^2\Gamma^2}}^2.
\end{equation}
For zero energy $E=0$ the above transmission is zero at $E_{1'}=-a^2E_{1}$.

\subsection{Second order von Neumann (2vN) approach}
For the interacting case, we use the second order von Neumann (2vN) approach,~\cite{PedersenPRB2005}
%\cite{PedersenPRB2005a}
which contains both coherent effects and cotunneling. However, the method fails around and below the Kondo temperature where strong correlations between the dot and the lead electrons appear. For our considered spinless system the Kondo temperature can be defined through the mapping to generalized Anderson model as discussed by Kashcheyevs \textit{et al.}~\cite{KashcheyevsPRB2007,Footnote} Numerically, the thermopower is found in an iterative procedure by varying the voltage until the current vanishes. The governing equations of the method and the solution procedure of those equations is presented in \appendixname~\ref{sec:2vN}. In the numerical calculations we use $k_{\mathrm{B}}T\sim\Gamma$ and a finite temperature difference $\Delta{T}=T_{\mr{L}}-T_{\mr{R}}$ (about an order of magnitude lower) in order to achieve convergent and reliable results. Furthermore,  we define the average temperature $\overline{T}=(T_{\mr{L}}+T_{\mr{R}})/2$ as a reference point. All the calculations in Figs.~\ref{fig3}-\ref{fig5} are done using the 2vN approach. For the non-interacting case $U=0$ the 2vN approach reproduces the exact transmission formalism results.~\cite{JinJCP2008,KarlstromJPhysAMathTheor2013} We explicitly checked that for present calculations the values $k_{\mathrm{B}}\Delta{T}=V=0.1\Gamma$ give the linear conductance $G$ and the linear thermopower $S$ on the scale of Figs.~\ref{fig3}-\ref{fig5} within the 2vN approach. Additionally, the value of bandwidth $D=20\Gamma$ is chosen large enough that the results in the considered parameter regime do not depend on $D$.~\cite{HaldanePRL1978}

\section{\label{sec:results}Results}

Our main focus is on the zeros of the thermopower $S$, which is a rather distinct spectroscopic feature, especially when there is a sign change of $S$. To emphasize the parameter values where the zeros appear we plot the square root of the absolute value of the thermopower $\sqrt{\abs{S}}$ and use a grayscale scheme where the zero value corresponds to white.

\subsection{\label{subsec:uzero}Non-interacting case $U=0$}

\begin{figure}[t]
\includegraphics[width=0.95\columnwidth]{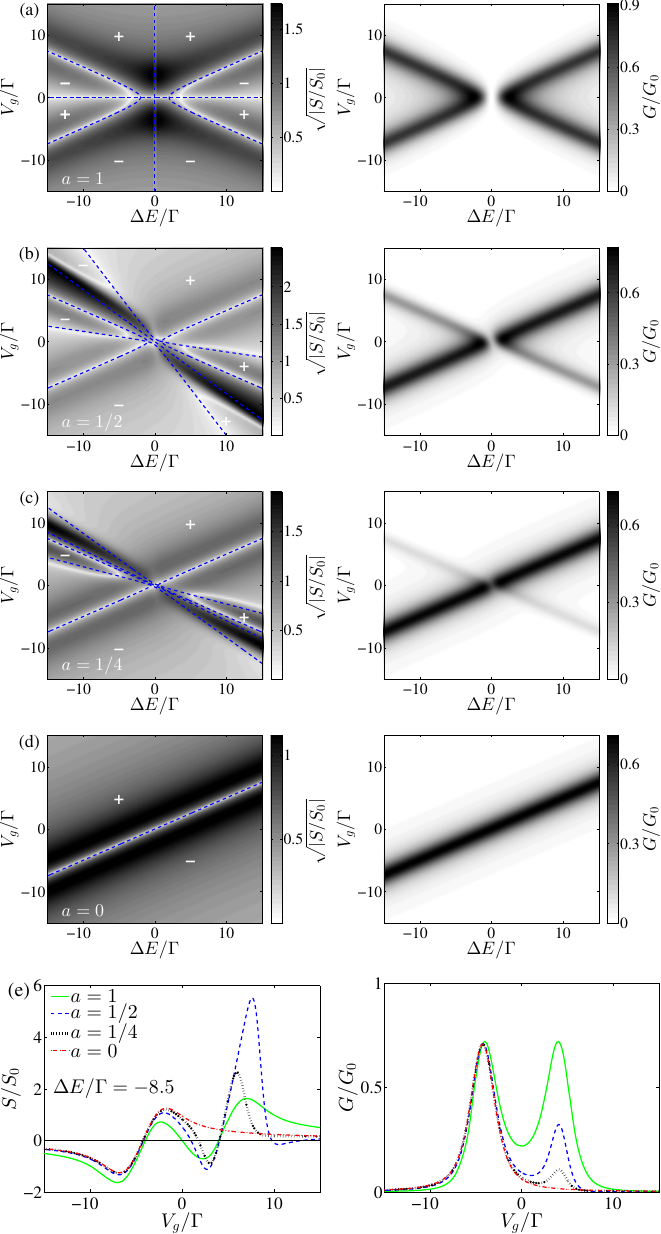}
\caption{\label{fig2} (Color online) The thermopower $S$ (left column) and conductance $G$ (right column) as a function of the detuning $\Delta{E}$ and gate voltage $V_{g}$ in the non-interacting case $U=0$. The asymmetry values are $a=1, \ 1/2, \ 1/4, \ 0$ for (a) to (d), respectively. For the contour plots the temperature is set to $k_{\mathrm{B}}T=0.5\Gamma$ and $\pm$ denotes the sign of the thermopower for these contour plots. The dashed (blue) curves depict zeros of thermopower, given by Eqs.~\eqref{anlzerostherm}, for the case of a small temperature $\Gamma\gg k_{\mathrm{B}}T\rightarrow 0$. Panel (e) gives a cut of the contour plots at the detuning $\Delta{E}=-8.5\Gamma$.}
\end{figure}

\begin{figure}[t]
\includegraphics[width=0.95\columnwidth]{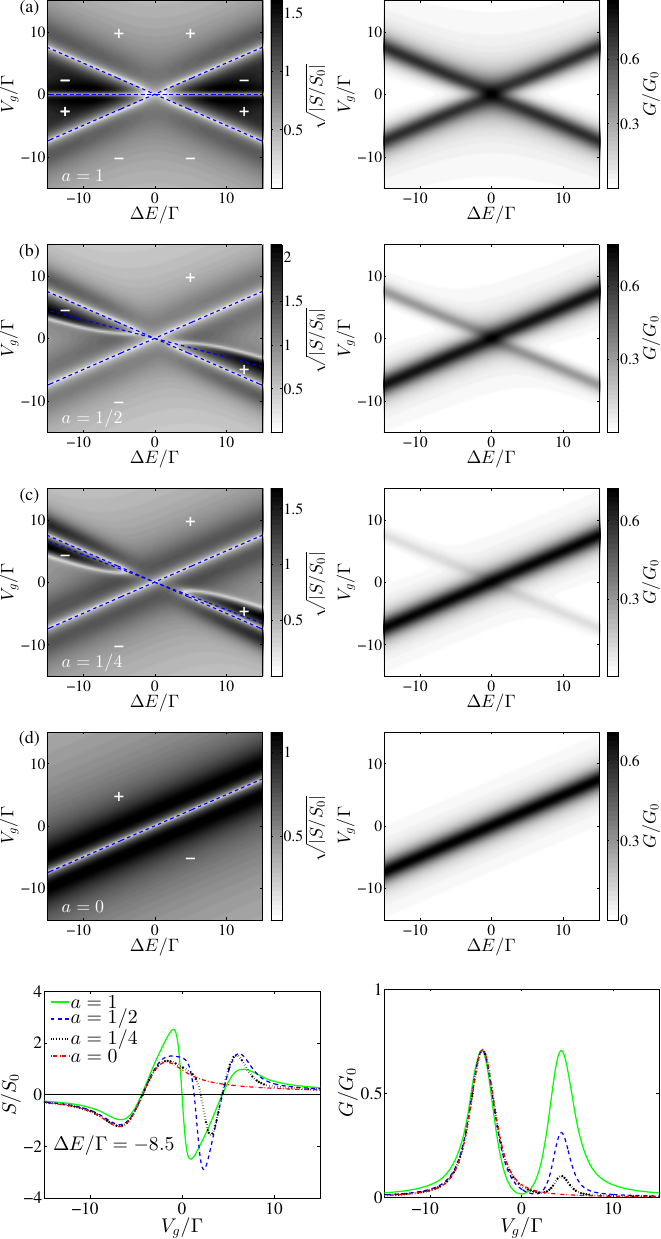} % Fig2S
\caption{\label{fig2S} (Color online) Same as \figurename~\ref{fig2} just for the case where the tunneling amplitudes have the same sign. The dashed (blue) curves depict zeros of thermopower, given by Eqs.~\eqref{SameSignZeros}.}
\end{figure}

We start by discussing the thermopower for the non-interacting $U=0$ case.\cite{NakanishiJPSJ2007,Xu-MingChinesePhysLett2009,ChoPRB2005,NakanishiJPSJ2007,Xu-MingChinesePhysLett2009,ZhengJLowTempPhys2012,GongJPSJ2012}  If the temperature is the lowest energy scale in the system, i.~e., $k_{\mathrm{B}}T\ll \Gamma$, the zeros of the thermopower can be obtained directly from the transmission function by using the Mott approximation,~\cite{MottBook1936,CutlerPR1969,AppleyardPRL1998,AppleyardPRB2000} which yields the following expression
\begin{equation}\label{TheMottFormula}
\frac{S_{M}}{S_{0}}\approx -\frac{\pi^2}{3}k_{\mr{B}}T\frac{\pd \ln{\mc{T}(E)}}{\pd E}\Bigg\rvert_{E=0}.
\end{equation}
Then we see that in this limit the zeros of thermopower are given by the zeros of the derivative of the transmission, i.~e., $S_{M}=0$, when $\pd_{E}\mc{T}(E)\rvert_{E=0}=0$. Using Eq.~\eqref{transanl}, this condition defines the following curves in the $(\Gamma,\Delta{E},V_{g})$ space:
\begin{subequations}\label{anlzerostherm}
\begin{align}
&\frac{\Delta{E}}{2}=\pm\sqrt{V_{g}^2+(a\Gamma)^2},\\
&\frac{\Delta{E}}{2}=-\frac{1-a^2}{1+a^2}V_{g},\\
&\frac{\Delta{E}}{2}(1-a^2)=-(1+a^2)V_{g}\pm a\sqrt{4V_{g}^2+[(1-a^2)\Gamma]^2}.
\end{align}
\end{subequations}
From the above expressions we see that for given detuning $\Delta{E}$ between the levels there can be up to five zeros, when the gate voltage $V_{g}$ is scanned. Three zeros appear once the distinct peaks, related to the energies of the single particle levels, become resolvable and this happens when the separation between the levels is larger than the energy scale determined by $\Gamma$. These zero's curves are given by the two solutions~(\ref{anlzerostherm}a), and by solution in~(\ref{anlzerostherm}c) with the minus sign for $V_{g}>0$ or the plus sign for $V_{g}<0$. Additional two zeros appear due to destructive interference, and for finite detuning $\Delta{E}\neq0$ this happens only when the couplings to the levels are asymmetric, $a\neq1$. For symmetric couplings, $a=1$, destructive interference leads to zero thermopower for $\Delta{E}=0$, but without the sign change, because it is a solution of double multiplicity. Figure~\ref{fig2} depicts the zeros given by analytical expressions Eqs.~\eqref{anlzerostherm} by blue dashed curves for different asymmetries $a$. These lines can be compared to the thermopower $\sqrt{\abs{S/S_0}}$ (in grayscale) where the temperature is comparable to the coupling $\Gamma$. We see that for larger values of $\Delta{E}$ or $V_{g}$ the zeros, corresponding to single particle levels, for the finite temperature case match Eqs.~\eqref{anlzerostherm}, but there are some discrepancies when $\Delta{E}$ and $V_{g}$ do not exceed $k_{\mathrm{B}}T$. For the interference features we see that they are sensitive to finite temperature, i.~e., depending on the coupling asymmetry they appear for larger values of $\Delta{E}, \ V_{g}\gg k_{\mathrm{B}}T$. \figurename~\ref{fig2}e shows the calculated data for a fixed $\Delta E=-8.5 \Gamma$. For $a=1/2$ (dashed blue curve) we clearly see the five  zeros in thermopower for the finite temperature calculation.

For the case of the same sign tunneling amplitudes from Eqs.~\eqref{transanlS} and~\eqref{TheMottFormula} we find that there can be up to three zeros in the thermopower for $T\rightarrow0$ for given $\Delta{E}$:
\begin{subequations}\label{SameSignZeros}
\begin{align}
&\frac{\Delta{E}}{2}=\pm V_{g},\\
&\label{SameSignZerosDest}\frac{\Delta{E}}{2}=-\frac{1+a^2}{1-a^2}V_{g}.
\end{align}
\end{subequations}
For low energies in the transmission Eq.~\eqref{transanlS} the destructive interference corresponds to a zero given by Eq.~\eqref{SameSignZerosDest}, i.e., the gate voltage has to be within the window of energies given by the detuning, i.e., $\abs{V_{g}}<\abs{\Delta{E}}$. This is in contrast to the canyon of conductance suppression situation, where the destructive interference appears for $\abs{V_g}>\abs{\Delta{E}}$. Additionally, the canyon of conductance suppression exhibits a thermopower zero for $\abs{V_{g}}<\abs{\Delta{E}}$ just because there is a minimum of conductance, and there is no necessity for destructive interference for this zero to appear. On the other hand, for the same sign amplitudes the large values of the thermopower close to a zero give a hint for the presence of the destructive interference (see \figurename~\ref{fig2S}).~\cite{NakanishiJPSJ2007}

The role of finite temperature can be understood as follows. In the different limit, when the tunneling coupling is the smallest energy scale, i.~e., $\Gamma\ll \abs{E-E_{1}}, \ \abs{E-E_{1'}}$, we obtain the following expression for transmission to lowest order in $\Gamma$
\begin{equation}\label{largeTtrans}
\mc{T}(E)\approx\pi\Gamma\left[\delta(E-E_1)+a^2\delta(E-E_{1'})\right],
\end{equation}
which is valid for both Eq.~\eqref{transanl} and Eq.~\eqref{transanlS}.
Applying Eq.~\eqref{uezS} yields the condition for zeros of thermopower
\begin{equation}\label{largeTeq}
\begin{aligned}
&x\cosh^{2}y+a^2y\cosh^{2}x=0,\\
&x=E_1/(2k_{\mathrm{B}}T), \quad y=E_{1'}/(2k_{\mathrm{B}}T).
\end{aligned}
\end{equation}
The above Eq.~\eqref{largeTeq} can have only up to three gate voltage $V_{g}$ solutions for given $\Delta{E}$, i.~e., the features arising due to destructive interference for the canyon of conductance suppression get washed out due to high temperature [as also seen from Eq.~\eqref{largeTtrans}]. Additionally, for symmetric coupling, $a=1$, Eq.~\eqref{largeTeq} yields the following condition for resolution of two levels,
\begin{equation}
\e^{r}=\frac{r+1}{r-1}, \quad r=\frac{\Delta{E}}{2k_{\mathrm{B}}T},
\end{equation}
which has a solution $\abs{\Delta{E}_{c}}\approx 3.09k_{\mathrm{B}} T$.~\cite{MeiravPRL1990} From \figurename~\ref{fig2}a we see that both the temperature $T$ and the coupling $\Gamma$ contribute to the effective broadening and the two levels become resolvable in the thermopower for $\Delta{E}\sim2\Gamma+3k_{\mathrm{B}}T$ in the case of the canyon of conductance suppression. For the same sign tunneling amplitudes we obtain that the resolution of the thermopower zeros around $\Delta{E}=0$ is determined just by the temperature broadening and not by $\Gamma$ (see \figurename~\ref{fig2S}a).

\begin{figure}[t]
\includegraphics[width=0.95\columnwidth]{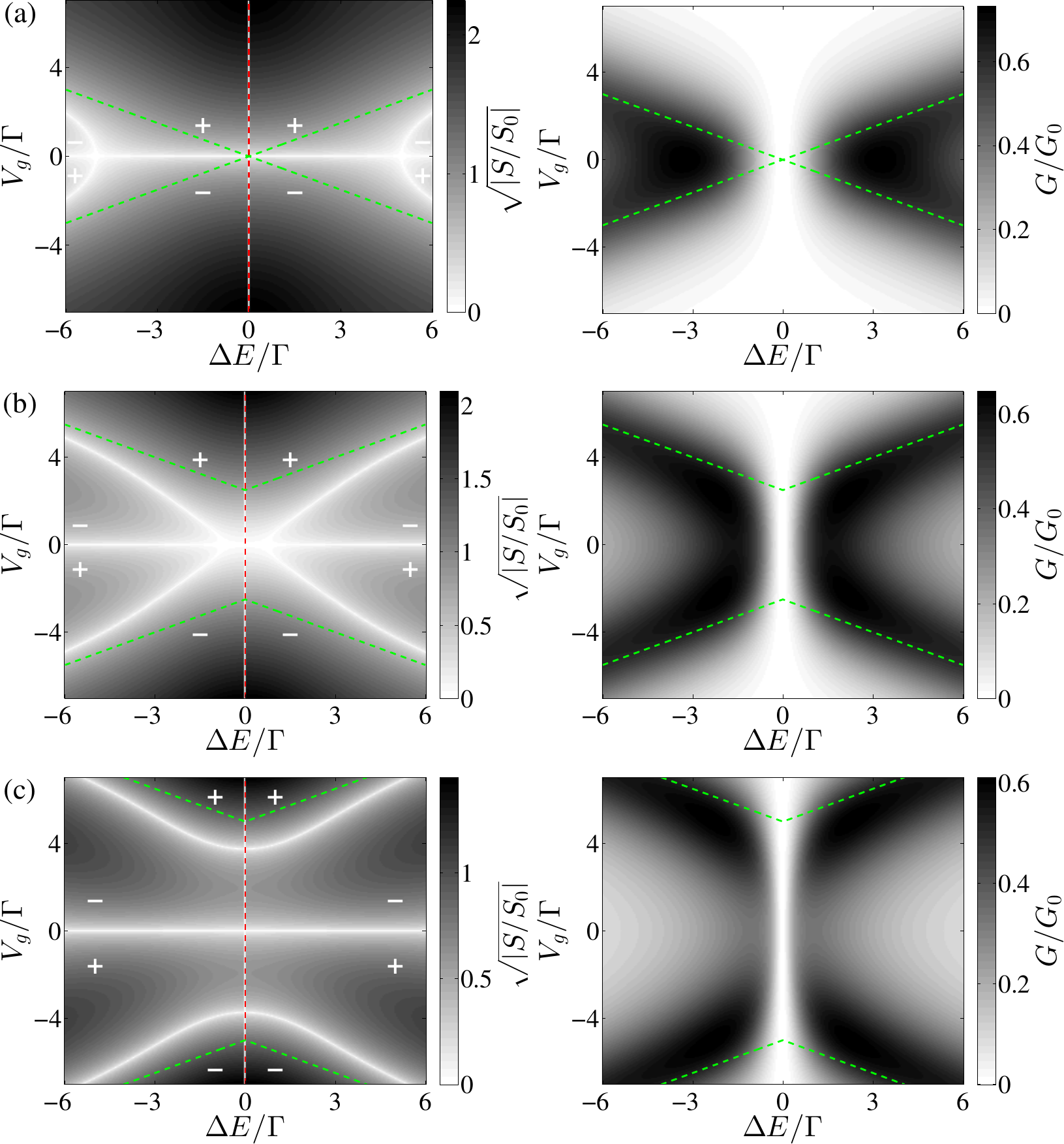} % Fig3
 \caption{\label{fig3} (Color online) The thermopower $S$ and conductance $G$ evolution as a function of increasing charging energy for symmetric coupling configuration $a=1$. The charging energy values are
 $U=0, \ 5\Gamma, \ 10\Gamma$
 for (a) to (c), respectively.
 For the 2vN thermopower calculations we have set
 $k_{\mathrm{B}}\Delta{T}=0.1\Gamma$,
 $k_{\mathrm{B}}\overline{T}=\Gamma$
 and for conductance calculations we have set $k_{\mathrm{B}}\Delta{T}=0$,
 $k_{\mathrm{B}}\overline{T}=\Gamma$,
 $V=0.1\Gamma$.
 Also a finite bandwidth
 $D=20\Gamma$
 is used. The dashed (green) lines denote the bare resonances given by Eq.~\eqref{trnen}. The dashed (red) lines show a region of detuning, where the thermopower is not defined because of the zero conductance. }
\end{figure}

\begin{figure}[!t]
\includegraphics[width=0.95\columnwidth]{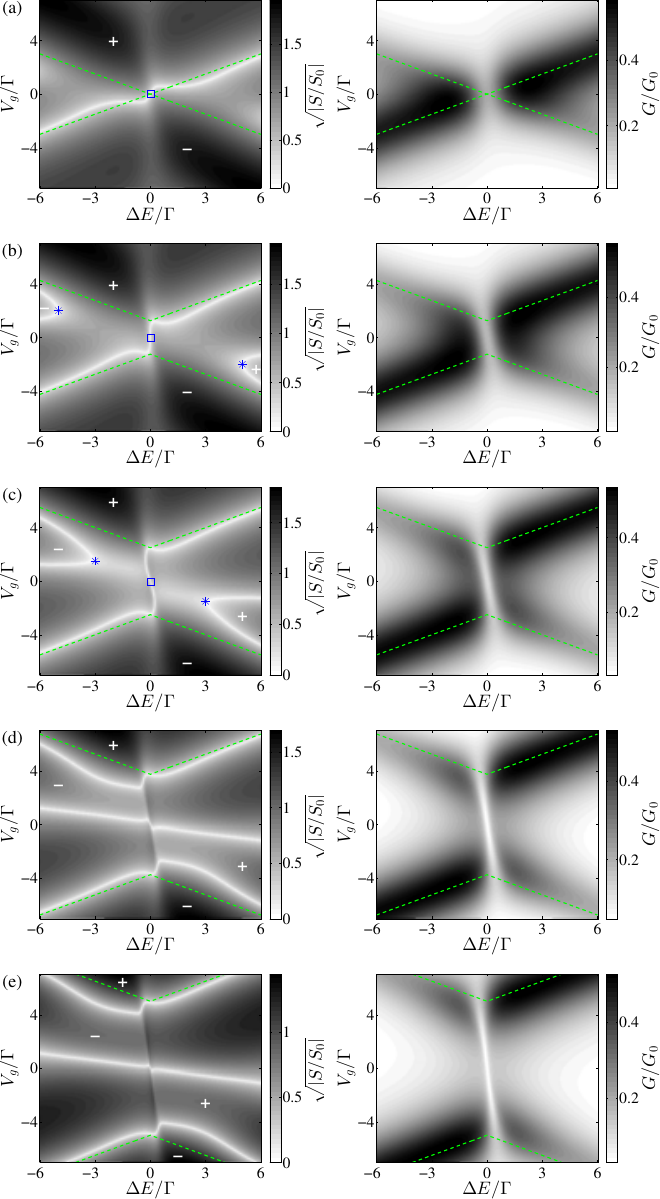} % Fig4
 \caption{\label{fig4}
 (Color online) The thermopower $S$ and conductance $G$ evolution as a function of increasing charging energy for asymmetric coupling configuration $a=1/2$. The charging energy values are
 $U=0, \ 2.5\Gamma, \ 5\Gamma, \ 7.5\Gamma, \ 10\Gamma$ for (a) to (e), respectively. Other parameter values are as in \figurename~\ref{fig3}.
 }
\end{figure}

\begin{figure}
\includegraphics[width=0.95\columnwidth]{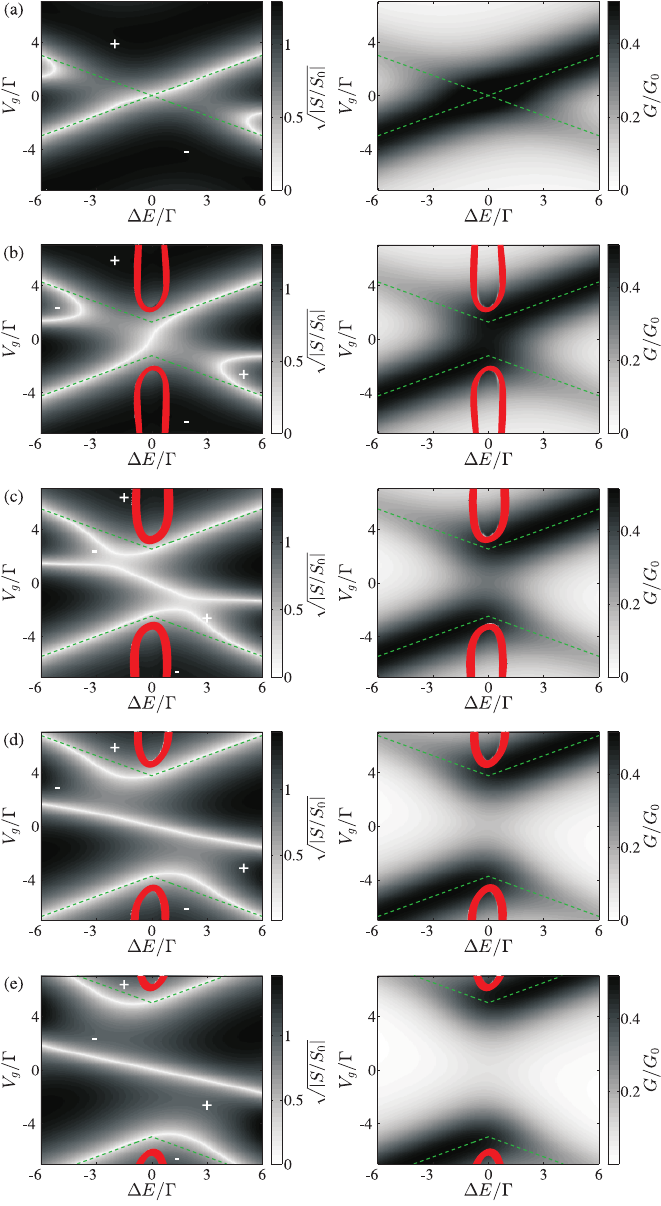} % Fig4S
 \caption{\label{fig4S}
 (Color online) The thermopower $S$ and conductance $G$ evolution as a function of increasing charging energy for the same sign tunneling amplitudes Eq.~\eqref{amplitudesS} and asymmetric coupling configuration $a=1/2$. The charging energy values are
 $U=0, \ 2.5\Gamma, \ 5\Gamma, \ 7.5\Gamma, 10\Gamma$ for (a) to (c), respectively. Other parameter values are as in \figurename~\ref{fig3}. The (red) horseshoe patches denote the region of parameters, where our numerical procedure for solving the 2vN equations does not converge.}
\end{figure}

\begin{figure}[t]
\includegraphics[width=0.95\columnwidth]{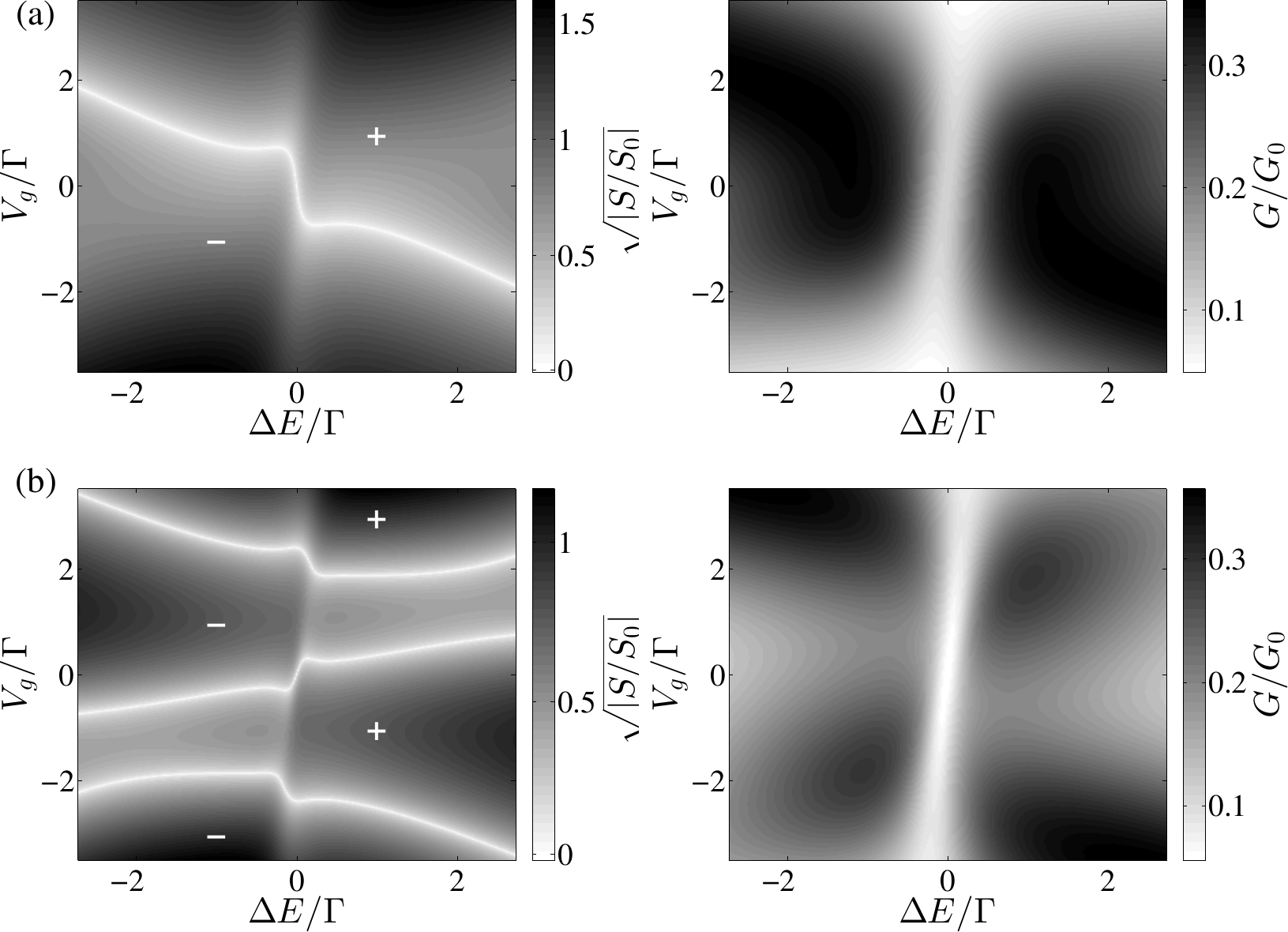}
 \caption{\label{fig5} The thermopower $S$ (first column) calculated for canyon of conductance $G$ suppression (second column) discussed in Ref.~\onlinecite{NilssonPRL2010}. The charging energy values are $U=2\Gamma$ and $U=5\Gamma$ for (a) and (b), respectively. Note that here the parametrization $t_{\mr{L}1}=-\sqrt{0.3}t_{0}$, $t_{\mr{R}1}=\sqrt{0.1}t_{0}$ $t_{\mr{L}1'}=t_{0}$, $t_{\mr{R}1'}=\sqrt{0.4}t_{0}$, with $t_{0}=\sqrt{\Gamma/(2\pi\nu_{F})}$, for tunneling amplitudes is used. Other parameter values are as in \figurename~\ref{fig3}.
 }
\end{figure}

\subsection{\label{subsec:ufinite}Finite $U\neq 0$ case}

In order to study the impact of the finite interaction  $U\neq0$, we first examine a case of equal coupling strengths for the two levels, i.~e., $a=1$. The corresponding thermopower and conductance are shown in \figurename~\ref{fig3}. Changing the interaction strength $U$ causes a minor change in the conductance, while the changes are more visible in the thermopower. We note that for $U\neq0$ the 2vN method can give negative diagonal reduced matrix elements $\Phi^{[0]}_{bb}$ defined in Eq.~\eqref{phieqs} and also the linear conductance $G$ can become negative, which is unphysical.~\cite{NilssonPRL2010} This is the reason that we do not address the interference features in the thermopower appearing at large gate voltages $V_{g}$.

The conductance plots in the right column of \figurename~\ref{fig3} show, that the conductance peaks for large $\abs{\Delta{E}}$ appear at particular resonances, where the quantum dot ground states differing by a single charge cross in energy. These are the common sequential tunneling peaks occurring at
\begin{equation}\label{trnen}
V_{g}=\pm\frac{U+\abs{\Delta{E}}}{2},
\end{equation}
which are shown by dashed green lines in \figurename~\ref{fig3}. For zero detuning $\Delta{E}$ these \textit{sequential tunneling} lines are broken by a canyon of conductance suppression.

Now we consider the thermopower in the left column of \figurename~\ref{fig3}. Around the sequential tunneling lines there is an equal amount of electron and hole tunneling for the given resonance and the average energy of tunneling particles becomes zero.\cite{TurekPRB2002} This provides a zero in the thermopower. As can be seen for large $\abs{\Delta{E}}$ there is actually a zero close to the dashed green lines, as expected. A further zero is occurring in between (at $V_g=0$ for the case $a=1$ considered here), where electron and hole tunneling for different resonances compensate each other.

For the charging energy, $U\lesssim U_{c}=\Gamma+3k_{\mathrm{B}}T$, smaller than the effective broadening of the levels, there is a region of detuning, where the level crossing of empty with singly occupied and singly with doubly occupied states in the dot can not be resolved (see \figurename~\ref{fig3}a,b). For the corresponding thermopower, there is only one sign shift for all gate voltages around zero detuning. For larger detunings $\Delta E\gtrsim U_{c}-U$ the levels can be resolved and again three sign shifts appear in the thermopower: at the two-level crossings and at the electron-hole symmetry point $V_{g}=0$. When the charging energy becomes larger than the effective broadening there are always three sign shifts in the thermopower as can bee seen from \figurename~\ref{fig3}c.

The behavior of thermopower and conductance for different $U$ when the coupling is asymmetric $a=1/2$ is shown in \figurename~\ref{fig4}. Here we focus on the thermopower zeros, which appear for low values of gate voltage and detuning, i.e., we do not address zeros, which appear due to interference (as discussed for $U=0$ above). The canyon of conductance suppression is now tilted due to the asymmetric coupling as discussed already in Ref.~\onlinecite{KarlstromPRB2011a}. More interestingly, the appearance of the thermopower zeros is changed quite drastically: for $U\lesssim U_c=5\Gamma$, the curve of zero thermopower becomes strongly tilted around $\Delta E=0$. It takes a characteristic S-shaped form, around its center denoted by a box in Figs.~\ref{fig4}a-c. Increasing $U$ pushes two further thermopower zeros (their onset is denoted by two asterisks in Figs.~\ref{fig4}b,c) towards the central S-shaped curve. Around critical $U_{c}$ there is a qualitative change (compare \figurename~\ref{fig4}c and \figurename~\ref{fig4}d), where the S-shaped curve merges with the other ones. Finally, for large $U$ (Fig.~\ref{fig4}e) the situation resembles the case for symmetric coupling, $a=1$, addressed in Fig.~\ref{fig3}c.

The situation when all the tunneling amplitudes have the same sign is depicted in \figurename~\ref{fig4S}. For $U>U_{c}$ we see that the evolution of zeros is qualitatively the same as in \figurename~\ref{fig4}. However, for the canyon of conductance suppression the zero lines have a jump around $\Delta{E}\approx 0$ (see Figs.~\ref{fig4}d and \ref{fig4}e). Additionally, for $U<U_{c}$ the S-shaped zero thermopower curve does not appear for the same sign configuration.

Lastly, in \figurename~\ref{fig5} we present the thermopower behavior for a system, whose conductance was studied experimentally in Ref.~\onlinecite{NilssonPRL2010}. These results may be relevant for future thermoelectric experiments on similar type devices, and shows the situation when the couplings have more general structure than Eq.~\eqref{amplitudes}. Note that for given parameters in \figurename~\ref{fig5} we have the level $E_{1'}$ more strongly coupled than the level $E_{1}$, which is opposite from the cases studied in Figs.~\ref{fig2}-\ref{fig4}. This results to a mirrored behavior of thermopower and conductance along zero detuning $\Delta{E}=0$. Otherwise, the introduced additional asymmetries in the couplings between the left and the right lead does not alter the qualitative behavior of the asymmetric coupling $a=1/2$ thermopower seen in \figurename~\ref{fig4}.

\section{\label{sec:concl}Conclusions}

The canyon of conductance suppression known from Refs.~\onlinecite{NilssonPRL2010,KarlstromPRB2011a} has been further investigated with thermopower $S$ acting as a probing tool. Zeros of the thermopower are a telling spectroscopic feature, which for the system studied in this paper yields information about the level coupling asymmetry $a$ (compare \figurename~\ref{fig3} vs.~\figurename~\ref{fig4}) and charging energy $U$ compared to the effective broadening of the levels given by $\sim 2\Gamma+3k_{\mathrm{B}}T$. Additionally, for the non-interacting case it was shown that up to five zeros can be observed when scanning the gate voltage for given detuning. This shows that the thermopower measurement could be useful to resolve features appearing due to destructive interference [see~\figurename~\ref{fig2} and Eq.~\eqref{anlzerostherm}]. Additionally, by comparing the same sign tunneling amplitude thermopower [Eq.~\eqref{amplitudes}] to canyon of conductance suppression thermopower it was shown that the canyon of conductance suppression has a unique signature around zero detuning $\Delta{E}=0$ (compare \figurename~\ref{fig4} vs.~\figurename~\ref{fig4S}).

\begin{acknowledgments}
We thank K.~G.~L.~Pedersen, A.~Donarini, and J.~K{\"{o}}nig for useful discussions. This work was supported by the Swedish Research Council (VR) and NanoLund.
\end{acknowledgments}

\appendix

\section{\label{sec:2vN}Second-order von Neumann (2vN) approach}
In the 2vN approach~\cite{PedersenPRB2005,PedersenPRB2007,PedersenPHE2010} we approximately solve the equation
\begin{equation}\label{vneq}
\i\hbar\frac{\pd}{\pd t}\rho=[H,\rho],
\end{equation}
by considering the density matrix $\rho$ elements, which connect the states differing by up to two electron or hole excitations. Such a treatment corresponds to the so-called resonant tunneling approximation in real-time diagrammatic approach,~\cite{KonigPRB1996} and yields an exact current for non-interacting systems ($U=0$ for our considered system).~\cite{KarlstromJPhysAMathTheor2013} We note that the 2vN method equations were originally derived in Ref.~\onlinecite{PedersenPRB2005}, and here we present this derivation for convenience with a slightly different notation. We write the governing equations in the many-particle eigenbasis $|a\rangle,|b\rangle,\ldots$ of the dot Hamiltonian $H_{\mr{D}}$~\eqref{hamD}. Expressed in this many-particle basis the tunneling Hamiltonian $H_{\mr{T}}$~\eqref{hamT} becomes
\begin{align}
\label{hamT2}
&H_{\mr{T}}=\sum_{ab,k\ell}\left(T_{ba,\ell}\ket{b}\bra{a}\can_{k\ell}+\mathrm{h.c.}\right),\\
&T_{ba,\ell}=\sum_{i=1,1'}t_{i\ell}\bra{b}\dd_{i}\ket{a}.
\end{align}
Here we used the \textit{letter convention}: if more than one state enters an equation, then the position of the letter in the alphabet follows the particle number (for example $N_b= N_a +1$, $N_c= N_a +2$, $N_{a'}=N_a$). In such a way the sum $\sum_{bc}$ restricts to those combinations, where $N_c = N_b+1$. For our considered system we have four many-particle eigenstates
\begin{equation}
\begin{aligned}
&\ket{0}, \quad &&E_{0}=0,\\
&\ket{1}=\dd_{1}\ket{0}, \quad &&E_{1}=-V_{g}+\frac{\Delta{E}}{2}-\frac{U}{2},\\
&\ket{1'}=\dd_{1'}\ket{0}, \quad &&E_{1'}=-V_{g}-\frac{\Delta{E}}{2}-\frac{U}{2},\\
&\ket{2}=\dd_{1'}\dd_{1}\ket{0}, \quad &&E_{2}=-2V_{g},
\end{aligned}
\end{equation}
and such many-particle tunneling amplitudes
\begin{equation}
\begin{aligned}
\bm{T}_{\ell}&=\begin{pmatrix}
0 & T_{01,\ell} &  T_{01',\ell} & 0 \\
T_{10,\ell} & 0 & 0 & T_{12,\ell} \\
T_{1'0,\ell} & 0 & 0 & T_{1'2,\ell} \\
0 & T_{21,\ell} & T_{21',\ell} & 0
\end{pmatrix}\\
&=\begin{pmatrix}
0 & t_{1\ell}^{*} &  t_{1'\ell}^{*} & 0 \\
t_{1\ell} & 0 & 0 & t_{1'\ell}^{*} \\
t_{1'\ell} & 0 & 0 & -t_{1\ell}^{*} \\
0 & t_{1'\ell} & -t_{1\ell} & 0
\end{pmatrix}.
\end{aligned}
\end{equation}

The density matrix elements are defined as
\begin{equation}
\rho_{ag,bg'}^{[n]}=\bra{ag}\rho\ket{bg'},
\end{equation}
where $\ket{bg}=\ket{b}\otimes\ket{g}$, with $\ket{b}$ denoting the eigenstate of the dot Hamiltonian~$H_{\mr{D}}$~\eqref{hamD} and $\ket{g}$ denoting the eigenstate of the lead Hamiltonian~$H_{\mr{LR}}$~\eqref{hamLR}. Here the label $n$ provides the number of electron or hole excitations needed to transform $\ket{g}$ into $\ket{g'}$. For example, we consider the matrix elements of the type
\begin{equation}
\begin{aligned}
&\rho_{bg,b'g}^{[0]}=\bra{bg}\rho\ket{b'g},\\
&\rho_{bg-\kappa,ag}^{[1]}=\bra{bg-\kappa}\rho\ket{ag},\\
&\rho_{dg-\kappa-\kappa',bg}^{[2]}=\bra{dg-\kappa-\kappa'}\rho\ket{bg},\\
&\rho_{bg-\kappa+\kappa',b'g}^{[2]}=\bra{bg-\kappa+\kappa'}\rho\ket{b'g}.
\end{aligned}
\end{equation}
Here we have introduced the following notation
\begin{equation}
\kappa\equiv k, \ \ell;
\end{equation}
\begin{equation}
\begin{aligned}
\ket{bg+\kappa}&=\ket{b}\otimes\cd_{\kappa}\ket{g},\\
\ket{bg-\kappa}&=\ket{b}\otimes\can_{\kappa}\ket{g},\\
\ket{dg-\kappa-\kappa'}&=\ket{d}\otimes\can_{\kappa'}\can_{\kappa}\ket{g},\\
\ket{bg-\kappa+\kappa'}&=\ket{b}\otimes\cd_{\kappa'}\can_{\kappa}\ket{g}.
\end{aligned}
\end{equation}
By neglecting all the density matrix elements with more than two electron or hole excitation $n>2$ from Eq.~\eqref{vneq} we obtain the equations
\begin{equation}\label{rheq0}
\begin{aligned}
\i\hbar\frac{\pd}{\pd{t}}\rho_{bg,b'g}^{[0]}&=(E_{b}-E_{b'})\rho_{bg,b'g}^{[0]}\\
&+\shs{\sum_{a_1,\kappa_1}}T_{ba_1,\kappa_1}\rho_{a_1g+\kappa_1,b'g}^{[1]}(-1)^{N_{a_1}}\\
&+\shs{\sum_{c_1,\kappa_1}}T_{bc_1,\kappa_1}\rho_{c_1g-\kappa_1,b'g}^{[1]}(-1)^{N_{b}}\\
&-\shs{\sum_{c_1,\kappa_1}}\rho_{bg,c_1g-\kappa_1}^{[1]}(-1)^{N_{b'}}T_{c_1b',\kappa_1}\\
&-\shs{\sum_{a_1,\kappa_1}}\rho_{bg,a_1g+\kappa_1}^{[1]}(-1)^{N_{a_1}}T_{a_1b',\kappa_1},
\end{aligned}
\end{equation}
\begin{equation}\label{rheq1}
\begin{aligned}
\i\hbar\frac{\pd}{\pd{t}}\rho_{cg-\kappa,bg}^{[1]}
&=(E_{c}-E_{\kappa}-E_{b})\rho_{cg-\kappa,bg}^{[1]}\\
&+\shs{\sum_{b_1,\kappa_1}}T_{cb_1,\kappa_1}\rho_{b_1g-\kappa+\kappa_1,bg}^{[2]}(-1)^{N_{b_1}}\\
&+\shs{\sum_{d_1,\kappa_1}}T_{cd_1,\kappa_1}\rho_{d_1g-\kappa-\kappa_1,bg}^{[2]}(-1)^{N_{c}}\\
&-\shs{\sum_{c_1,\kappa_1}}\rho_{cg-\kappa,c_1g-\kappa_1}^{[2]}(-1)^{N_{b}}T_{c_1b,\kappa_1}\\
&-\shs{\sum_{a_1,\kappa_1}}\rho_{cg-\kappa,a_1g+\kappa_1}^{[2]}(-1)^{N_{a_1}}T_{a_1b,\kappa_1},
%%%%%%%%%%%
%&=(E_{c}-E_{\kappa}-E_{b})\rho_{cg-\kappa,bg}^{[1]}\\
%&+\shs{\sum_{b_1,\kappa_1}}T_{cb_1,\kappa}\rho_{b_1g,bg}^{[0]}(-1)^{N_{b_1}}\bra{g}\cd_{\kappa}\can_{\kappa}\ket{g}\\
%&-\shs{\sum_{c_1,\kappa_1}}\rho_{cg-\kappa,c_1g-\kappa}^{[0]}(-1)^{N_{b}}T_{c_1b,\kappa}.
\end{aligned}
\end{equation}
\begin{equation}\label{rheq2a}
\begin{aligned}
\i\hbar\frac{\pd}{\pd{t}}\rho_{bg-\kappa+\kappa',b'g}^{[2]}
&\approx(E_{b}-E_{\kappa}+E_{\kappa'}-E_{b'})\rho_{bg-\kappa+\kappa',b'g}^{[2]}\\
&+\shs{\sum_{a_1}}T_{ba_1,\kappa}\rho_{a_1g-\kappa+\kappa'+\kappa,b'g}^{[1]}(-1)^{N_{a_1}}\\%\bra{g}\cd_{\kappa}\can_{\kappa}\ket{g}\\
&+\shs{\sum_{c_1}}T_{bc_1,\kappa'}\rho_{c_1g-\kappa+\kappa'-\kappa',b'g}^{[1]}(-1)^{N_{b}}\\%\bra{g}\can_{\kappa'}\cd_{\kappa'}\ket{g}\\
&-\shs{\sum_{c_1}}\rho_{bg-\kappa+\kappa',c_1g-\kappa}^{[1]}(-1)^{N_{b'}}T_{c_1b',\kappa}\\
&-\shs{\sum_{a_1}}\rho_{bg-\kappa+\kappa',a_1g+\kappa'}^{[1]}(-1)^{N_{a_1}}T_{a_1b',\kappa'},
\end{aligned}
\end{equation}
\begin{equation}\label{rheq2b}
\begin{aligned}
\i\hbar\frac{\pd}{\pd{t}}\rho_{dg-\kappa+\kappa',bg}^{[2]}
&\approx(E_{d}-E_{\kappa}-E_{\kappa'}-E_{b})\rho_{dg-\kappa-\kappa',bg}^{[2]}\\
&+\shs{\sum_{c_1}}T_{dc_1,\kappa}\rho_{c_1g-\kappa-\kappa'+\kappa,bg}^{[1]}(-1)^{N_{c_1}}\\%\bra{g}\cd_{\kappa}\can_{\kappa}\ket{g}\\%\delta_{\kappa,1}
&+\shs{\sum_{c_1}}T_{dc_1,\kappa'}\rho_{c_1g-\kappa-\kappa'+\kappa',bg}^{[1]}(-1)^{N_{c_1}}\\%\bra{g}\cd_{\kappa'}\can_{\kappa'}\ket{g}\\%\delta_{\kappa',1}
&-\shs{\sum_{c_1}}\rho_{dg-\kappa-\kappa',c_1g-\kappa}^{[1]}(-1)^{N_{b}}T_{c_1b,\kappa}\\
&-\shs{\sum_{c_1}}\rho_{dg-\kappa-\kappa',c_1g-\kappa'}^{[1]}(-1)^{N_{b}}T_{c_1b,\kappa'}.
\end{aligned}
\end{equation}
Note that all indices with subscript 1 like $a_1, \ c_1, \ \kappa_1$ are summed over. Additionally, phase factors like $(-1)^{N_{b}}$ appear due to order exchange of the lead operators with the dot operators, i.e., $\can_{\kappa}(\ket{b}\otimes\ket{g})=(-1)^{N_{b}}\ket{b}\otimes\can_{\kappa}\ket{g}$.

\vspace{6cm}

Summing Eqs.~\eqref{rheq0} and \eqref{rheq1} over all the lead states $\ket{g}$ we get
\begin{equation}\label{pheq0}
\begin{aligned}
\i\hbar\frac{\pd}{\pd{t}}\Phi_{bb'}^{[0]}&=(E_{b}-E_{b'})\Phi_{bb'}^{[0]}\\
&+\shs{\sum_{a_1,\kappa_1}}T_{ba_1,\kappa_1}\Phi_{a_1b',\kappa_1}^{[1]}
+\shs{\sum_{c_1,\kappa_1}}T_{bc_1,\kappa_1}\Phi_{c_1b',\kappa_1}^{[1]}\\
&-\shs{\sum_{c_1,\kappa_1}}\Phi_{bc_1,\kappa_1}^{[1]}T_{c_1b',\kappa_1}
-\shs{\sum_{a_1,\kappa_1}}\Phi_{ba_1,\kappa_1}^{[1]}T_{a_1b',\kappa_1},
\end{aligned}
\end{equation}
\begin{equation}\label{pheq1}
\begin{aligned}
\i\hbar\frac{\pd}{\pd{t}}\Phi_{cb,\kappa}^{[1]}&\approx(E_{c}-E_{\kappa}-E_{b})\Phi_{cb,\kappa}^{[1]}\\
&+\shs{\sum_{b_1}}T_{cb_1,\kappa}\Phi_{b_1b}^{[0]}f_{\kappa}
-\shs{\sum_{c_1}}\Phi_{cc_1}^{[0]}T_{c_1b,\kappa}f_{-\kappa},\\
&+\shs{\sum_{b_1,\kappa_1}}T_{cb_1,\kappa_1}\Phi_{b_1b,-\kappa+\kappa_1}^{[2]}
 +\shs{\sum_{d_1,\kappa_1}}T_{cd_1,\kappa_1}\Phi_{d_1b,-\kappa-\kappa_1}^{[2]}\\
&+\shs{\sum_{c_1,\kappa_1}}\Phi_{cc_1,-\kappa+\kappa_1}^{[2]}T_{c_1b,\kappa_1}
 +\shs{\sum_{a_1,\kappa_1}}\Phi_{ca_1,-\kappa-\kappa_1}^{[2]}T_{a_1b,\kappa_1},
\end{aligned}
\end{equation}
\begin{equation}\label{pheq2a}
\begin{aligned}
\i\hbar\frac{\pd}{\pd{t}}\Phi_{bb',-\kappa+\kappa'}^{[2]}&\approx(E_{b}-E_{\kappa}+E_{\kappa'}-E_{b'})\Phi_{bb',-\kappa+\kappa'}^{[2]}\\
&-\shs{\sum_{a_1}}T_{ba_1,\kappa}\Phi_{a_1b',\kappa'}^{[1]}f_{\kappa}
 +\shs{\sum_{c_1}}T_{bc_1,\kappa'}\Phi_{c_1b',\kappa}^{[1]}f_{-\kappa'}\\
&-\shs{\sum_{c_1}}\Phi_{bc_1,\kappa'}^{[1]}T_{c_1b',\kappa}f_{-\kappa}
 +\shs{\sum_{a_1}}\Phi_{ba_1,\kappa}^{[1]}T_{a_1b',\kappa'}f_{\kappa'},
\end{aligned}
\end{equation}
\begin{equation}\label{pheq2b}
\begin{aligned}
\i\hbar\frac{\pd}{\pd{t}}\Phi_{db,-\kappa-\kappa'}^{[2]}&\approx(E_{d}-E_{\kappa}-E_{\kappa'}-E_{b})\Phi_{db,-\kappa-\kappa'}^{[2]}\\
&-\shs{\sum_{c_1}}T_{dc_1,\kappa}\Phi_{c_1b,\kappa'}^{[1]}f_{\kappa}
 +\shs{\sum_{c_1}}T_{dc_1,\kappa'}\Phi_{c_1b,\kappa}^{[1]}f_{\kappa'}\\
&-\shs{\sum_{c_1}}\Phi_{dc_1,\kappa'}^{[1]}T_{c_1b,\kappa}f_{-\kappa}
 +\shs{\sum_{c_1}}\Phi_{dc_1,\kappa}^{[1]}T_{c_1b,\kappa'}f_{-\kappa'}.
\end{aligned}
\end{equation}
where we introduced the following notation
\begin{align}
&\label{phieqs}\Phi_{bb'}^{[0]}=\sum_{g}\rho_{bg,b'g}^{[0]},\\
&\Phi_{cb,\kappa}^{[1]}=\sum_{g}\rho_{cg-\kappa,bg}^{[1]}(-1)^{N_{b}},\quad
 \Phi_{bc,\kappa}^{[1]}=\big[\Phi_{cb,\kappa}^{[1]}\big]^{*},\nonumber\\
&\Phi_{ca,-\kappa-\kappa'}^{[2]}=-\sum_{g}\rho_{cg-\kappa-\kappa',ag}^{[2]},\nonumber\\
&\Phi_{bb',-\kappa+\kappa'}^{[2]}=+\sum_{g}(1-\delta_{\kappa\kappa'})\rho_{bg-\kappa+\kappa',b'g}^{[2]},\nonumber\\
&f_{\kappa}\equiv f_{k\ell}=(\exp[(E_{k}-\mu_{\ell})/k_{\mr{B}}T_{\ell}]+1)^{-1},\nonumber\\
&f_{-\kappa}\equiv 1-f_{k\ell},\nonumber
\end{align}

\noindent and when going from Eq.~\eqref{rheq1} to Eq.~\eqref{pheq1} we have used

\begin{equation}
\begin{aligned}
%\bra{g}\cd_{\kappa}\can_{\kappa}\ket{g}
&\rho_{b_1g-\kappa+\kappa_1,bg}^{[2]}=\rho_{b_1g-\kappa+\kappa,bg}^{[0]}+(1-\delta_{\kappa\kappa_1})\rho_{b_1g-\kappa+\kappa_1,bg}^{[2]},\\
&\rho_{cg-\kappa,c_1g-\kappa_1}^{[2]}=\rho_{cg-\kappa,c_1g-\kappa}^{[0]}+(1-\delta_{\kappa\kappa_1})\rho_{cg-\kappa,c_1g-\kappa_1}^{[2]}.
\end{aligned}
\end{equation}

\noindent Here we have also assumed that the electrons in the leads are \textit{thermally distributed} according to the Fermi-Dirac distribution $f$ and that this distribution is not affected by the coupling to the quantum dots. This assumption leads to the following relations for Eq.~\eqref{rheq1}
\begin{equation}
\begin{aligned}
&\sum_{g}\rho_{b_1g-\kappa+\kappa,bg}^{[0]}\approx f_{\kappa}\Phi_{b_1b}^{[0]},\\
&\sum_{g}\rho_{cg-\kappa,c_1g-\kappa}^{[0]}\approx f_{-\kappa}\Phi_{cc_1}^{[0]},
\end{aligned}
\end{equation}
\noindent for Eq.~\eqref{rheq2a}

\begin{align}
&\sum_{g}\rho_{a_1g-\kappa+\kappa'+\kappa,b'g}^{[1]}(-1)^{N_{a_1}}\approx-f_{\kappa}\Phi_{a_1b',\kappa'}^{[1]},\nonumber\\
&\sum_{g}\rho_{c_1g-\kappa+\kappa'-\kappa',b'g}^{[1]}(-1)^{N_{b}}\approx f_{-\kappa'}\Phi_{c_1b',\kappa}^{[1]},\\
&\sum_{g}\rho_{bg-\kappa+\kappa',c_1g-\kappa}^{[1]}(-1)^{N_{b'}}\approx f_{-\kappa}\Phi_{bc_1,\kappa'}^{[1]},\nonumber\\
&\sum_{g}\rho_{bg-\kappa+\kappa',a_1g+\kappa'}^{[1]}(-1)^{N_{a_1}}\approx-f_{\kappa'}\Phi_{ba_1,\kappa}^{[1]},\nonumber
\end{align}

\noindent and for Eq.~\eqref{rheq2b}

\begin{align}
&\sum_{g}\rho_{c_1g-\kappa-\kappa'+\kappa,bg}^{[1]}(-1)^{N_{c_1}}\approx-\Phi_{c_1b,\kappa'}^{[1]}f_{\kappa},\nonumber\\
&\sum_{g}\rho_{c_1g-\kappa-\kappa'+\kappa',bg}^{[1]}(-1)^{N_{c_1}}\approx\Phi_{c_1b,\kappa}^{[1]}f_{\kappa'},\\
&\sum_{g}\rho_{dg-\kappa-\kappa',c_1g-\kappa}^{[1]}(-1)^{N_{b}}\approx\Phi_{dc_1,\kappa'}^{[1]}f_{-\kappa},\nonumber\\
&\sum_{g}\rho_{dg-\kappa-\kappa',c_1g-\kappa'}^{[1]}(-1)^{N_{b}}\approx-\Phi_{dc_1,\kappa}^{[1]}f_{-\kappa'}.\nonumber
\end{align}

\begin{widetext}
For the stationary state we assume the conditions
\begin{equation}\label{statcond}
\begin{aligned}
\i\hbar\frac{\pd}{\pd{t}}\Phi_{bb'}^{[0]}=0,\quad
\i\hbar\frac{\pd}{\pd{t}}\Phi_{cb,\kappa}^{[1]}=0,\quad
\i\hbar\frac{\pd}{\pd{t}}\Phi_{bb',-\kappa+\kappa'}^{[2]}=0,\quad
\i\hbar\frac{\pd}{\pd{t}}\Phi_{db,-\kappa-\kappa'}^{[2]}=0,
\end{aligned}
\end{equation}
which allow to write $\Phi^{[2]}$ in terms of $\Phi^{[1]}$ as
\begin{equation}\label{ssphi2_2vN}
\begin{aligned}
&\Phi_{bb',-\kappa+\kappa'}^{[2]}=
\frac{-\shs{\sum_{a_1}}T_{ba_1,\kappa}\Phi_{a_1b',\kappa'}^{[1]}f_{\kappa}
 +\shs{\sum_{c_1}}T_{bc_1,\kappa'}\Phi_{c_1b',\kappa}^{[1]}f_{-\kappa'}
-\shs{\sum_{c_1}}\Phi_{bc_1,\kappa'}^{[1]}T_{c_1b',\kappa}f_{-\kappa}
 +\shs{\sum_{a_1}}\Phi_{ba_1,\kappa}^{[1]}T_{a_1b',\kappa'}f_{\kappa'}}{E_{\kappa}-E_{\kappa'}-E_{b}+E_{b'}+\i\eta},\\
%%%%%%%%%%%%%%%%%%%
&\Phi_{db,-\kappa-\kappa'}^{[2]}=
\frac{-\shs{\sum_{c_1}}T_{dc_1,\kappa}\Phi_{c_1b,\kappa'}^{[1]}f_{\kappa}
 +\shs{\sum_{c_1}}T_{dc_1,\kappa'}\Phi_{c_1b,\kappa}^{[1]}f_{\kappa'}
-\shs{\sum_{c_1}}\Phi_{dc_1,\kappa'}^{[1]}T_{c_1b,\kappa}f_{-\kappa}
 +\shs{\sum_{c_1}}\Phi_{dc_1,\kappa}^{[1]}T_{c_1b,\kappa'}f_{-\kappa'}}{E_{\kappa}+E_{\kappa'}-E_{d}+E_{b}+\i\eta}.
\end{aligned}
\end{equation}
Here we have added a positive infinitesimal $\eta=+0$ to ensure a proper decay of initial conditions. After inserting the above expressions into Eq.~\eqref{pheq1} we obtain the integral equations of the 2vN method for the stationary state
\begin{subequations}
\begin{equation}\label{eq2vN}
\begin{aligned}
%%%%%%%%%%
0&=-(E_{\kap}-E_{c}+E_{b}+\i\eta)\Phi^{[1]}_{cb,\kap}
+T_{cb_1,\kap}f_{\kap}\Phi^{[0]}_{b_1b}
-\Phi^{[0]}_{cc_1}f_{-\kap}T_{c_1b,\kap}\\
%1
&%\phantom{...}
+\frac{T_{cb_1,\kap_1}\left[
\color{black}{
T_{b_1c_1,\kap_1}f_{-\kap_1}\Phi^{[1]}_{c_1b,\kap}
+\Phi^{[1]}_{b_1a_1,\kap}f_{\kap_1}T_{a_1b,\kap_1}}
\color{black}{
-T_{b_1a_1,\kap}f_{\kap}\Phi^{[1]}_{a_1b,\kap_1}
-\Phi^{[1]}_{b_1c_1,\kap_1}f_{-\kap}T_{c_1b,\kap}}
\right]}{E_{\kap}-E_{\kap_1}-E_{b_1}+E_{b}+\i\eta}\\
%2
&%\phantom{...}
+\frac{T_{cd_1,\kap_1}\left[
\color{black}{
T_{d_1c_1,\kap_1}f_{\kap_1}\Phi^{[1]}_{c_1b,\kap}
+\Phi^{[1]}_{d_1c_1,\kap}f_{-\kap_1}T_{c_1b,\kap_1}}
\color{black}{
-T_{d_1c_1,\kap}f_{\kap}\Phi^{[1]}_{c_1b,\kap_1}
-\Phi^{[1]}_{d_1c_1,\kap_1}f_{-\kap}T_{c_1b,\kap}}
\right]}{E_{\kap}+E_{\kap_1}-E_{d_1}+E_{b}+\i\eta}\\
%3
&%\phantom{...}
+\frac{\left[
\color{black}{
T_{cd_1,\kap_1}f_{-\kap_1}\Phi^{[1]}_{d_1c_1,\kap}
+\Phi^{[1]}_{cb_1,\kap}f_{\kap_1}T_{b_1c_1,\kap_1}}
\color{black}{
-T_{cb_1,\kap}f_{\kap}\Phi^{[1]}_{b_1c_1,\kap_1}
-\Phi^{[1]}_{cd_1,\kap_1}f_{-\kap}T_{d_1c_1,\kap}}
\right]T_{c_1b,\kap_1}}{E_{\kap}-E_{\kap_1}-E_{c}+E_{c_1}+\i\eta}\\
%4
&%\phantom{...}
+\frac{\left[
\color{black}{
T_{cb_1,\kap_1}f_{\kap_1}\Phi^{[1]}_{b_1a_1,\kap}
+\Phi^{[1]}_{cb_1,\kap}f_{-\kap_1}T_{b_1a_1,\kap_1}}
\color{black}{
-T_{cb_1,\kap}f_{\kap}\Phi^{[1]}_{b_1a_1,\kap_1}
-\Phi^{[1]}_{cb_1,\kap_1}f_{-\kap}T_{b_1a_1,\kap}}
\right]T_{a_1b,\kap_1}}{E_{\kap}+E_{\kap_1}-E_{c}+E_{a_1}+\i\eta},
%%%%%%%%%%
\end{aligned}
\end{equation}

\begin{equation}
0=(E_{b}-E_{b'})\Phi_{bb'}^{[0]}
+\shs{\sum_{a_1,\kappa_1}}T_{ba_1,\kappa_1}\Phi_{a_1b',\kappa_1}^{[1]}
+\shs{\sum_{c_1,\kappa_1}}T_{bc_1,\kappa_1}\Phi_{c_1b',\kappa_1}^{[1]}
-\shs{\sum_{c_1,\kappa_1}}\Phi_{bc_1,\kappa_1}^{[1]}T_{c_1b',\kappa_1}
-\shs{\sum_{a_1,\kappa_1}}\Phi_{ba_1,\kappa_1}^{[1]}T_{a_1b',\kappa_1}.\quad\quad
\end{equation}
\end{subequations}
\end{widetext}
Additionally, we impose the normalisation condition for the diagonal reduced-density matrix elements:
\begin{equation}
\sum_{b}\Phi_{bb}^{[0]}=1.
\end{equation}
The integral equation \eqref{eq2vN} under interest has the structure
\begin{equation}
\Phi^{[1]}_{\kappa}=F_{\kappa}+\sum_{\kappa_1}K_{\kappa,\kappa_1}\Phi^{[1]}_{\kappa_1}.
\end{equation}
It is solved iteratively on an equidistant energy grid $E_{k}$ by having $N=2^{13}$ discretization points for our considered calculations. The zeroth iteration of $\Phi^{[1]}_{\kappa,0}=F_{\kappa}$ is determined by making a local approximation, i.e., terms of the form $\Phi^{[1]}_{ba,\kappa_1}$ which have integrated momentum label $\kappa_1$ are neglected. Then the first correction is determined as $\delta\Phi^{[1]}_{\kappa,1}=\sum_{\kappa_1}K_{\kappa,\kappa_1}F_{\kappa_1}$. The higher order corrections are given by $\delta\Phi^{[1]}_{\kappa,n}=\sum_{\kappa_1}K_{\kappa,\kappa_1}\delta\Phi^{[1]}_{\kappa',n-1}$ and the solution is expressed as $\Phi^{[1]}_{\kappa}=\Phi^{[1]}_{\kappa,0}+\sum_{n}\delta\Phi^{[1]}_{\kappa',n}$. We make iterations up to $n=6$, which yields good convergence for most parameter values. In these iterations we need to evaluate a lot Hilbert transform of the form:
\begin{equation}
\begin{aligned}
&H(\Phi^{[1]}_{\kappa})=\frac{1}{\pi}\int_{-D}^{D}\frac{\Phi^{[1]}_{\kappa'}\dif{E_{\kappa'}}}{E_{\kappa}-E_{\kappa'}\pm\i\eta}\\
&\phantom{.......}
=\frac{1}{\pi}\mc{P}\int_{-D}^{D}\frac{\Phi^{[1]}_{\kappa'}\dif{E_{\kappa'}}}{E_{\kappa}-E_{\kappa'}}\mp\i\Phi^{[1]}_{\kappa}\theta(D-\abs{E_{\kappa}}).
\end{aligned}
\end{equation}
The principal value integrals are efficiently evaluated on equidistant grid with $N$ points using fast Fourier transform, which has complexity $O(N\log{N})$.~\cite{FrederiksenMaster2004,PressBook2007}

Finally, we are interested in the current going from the lead $\ell$ into the quantum dots, which is given by
\begin{equation}\label{cureq}
I_{\ell}(t)=\frac{2e}{\hbar}\sum_{k}\Imag[T_{bc,k\ell}\Phi_{cb,k\ell}^{[1]}],
\end{equation}
and which shows that $\Phi_{cb,k\ell}^{[1]}$ are the energy resolved current amplitudes.

\section{\label{sec:mapAnderson}Mapping to generalized Anderson model}

In linear response to applied bias $V$ or temperature difference $\Delta{T}$ the Hamiltonian Eq.~\eqref{hamFull} can be mapped to generalized Anderson model by mixing the left lead electrons with the right lead electrons, and the dot orbital $1$ with the orbital $1'$. The generalized Anderson model has the form:~\cite{KashcheyevsPRB2007}
\begin{equation}\label{genAnderson}
\begin{aligned}
&H=\sum_{k,\sigma}E_{k}^{\ph{\dag}}\cd_{k\sigma}\can_{k\sigma}
+\sum_{k,\sigma}V_{\sigma}(\cd_{k\sigma}\dan_{\sigma}+\mathrm{h.c.})+Un_{\up}n_{\down}\\
&+\sum_{\sigma}\left(\ve-\sigma\frac{h}{2}\cos\theta\right)n_{\sigma}
-(\dd_{\up}\dan_{\down}+\dd_{\down}\dan_{\up})\frac{h}{2}\sin\theta,
\end{aligned}
\end{equation}
where $n_{\sigma}=\cd_{k\sigma}\can_{k\sigma}$, $\sigma\in\{\up,\down\}$ represents a pseudo-spin and $h$ is an effective magnetic field. The Hamiltonian \eqref{hamFull} is expressed in the form of \eqref{genAnderson} by performing a singular value decomposition (SVD) on the tunneling Hamiltonian \eqref{hamT} as
\begin{equation}\label{hamttrans}
H_{\mr{T}}=\sum_{k}t
\begin{bmatrix}
\cd_{\mr{L}k} & \cd_{\mr{R}k}
\end{bmatrix}
\mc{A}
\begin{bmatrix}
\dan_{1} \\ \dan_{1'}
\end{bmatrix}
+\mr{h.c.},
\end{equation}
\begin{equation}
\mc{A}=
\begin{pmatrix}
t_{L1} & t_{L1'} \\
t_{R1} & t_{R1'}
\end{pmatrix}=
R_{l}^{\dag}
\begin{pmatrix}
V_{\up} & 0 \\
0 & V_{\down}
\end{pmatrix}R_{d}.
\end{equation}
Using the rotations $R_{l}$ and $R_{d}$ the pseudo-spin operators are expressed in terms of original basis operators as:
\begin{equation}
\begin{bmatrix}
c_{\up k} \\ c_{\down k}
\end{bmatrix}=
R_{l}\begin{bmatrix}
c_{\mr{L}k} \\ c_{\mr{R}k}
\end{bmatrix},
\quad
\begin{bmatrix}
d_{\up} \\ d_{\down}
\end{bmatrix}=
R_{d}\begin{bmatrix}
d_{1} \\ d_{1'}
\end{bmatrix}.
\end{equation}
For the tunneling amplitudes of the form Eq.~\eqref{amplitudes} we obtain
\begin{align}
&\mc{A}_{-}=t
\begin{pmatrix}
1 & -a \\
1 & a
\end{pmatrix}=
R_{l}^{\dag}
t\begin{pmatrix}
\sqrt{2} & 0 \\
0 & \sqrt{2}a
\end{pmatrix}R_{d,-},\\
%\end{equation}
%
%\begin{equation*}
&R_{l}=\frac{1}{\sqrt{2}}
\begin{pmatrix}
+1 & +1 \\
-1 & +1
\end{pmatrix},\quad
R_{d,-}=\begin{pmatrix}
1 & 0 \\
0 & 1
\end{pmatrix},
\quad \theta_{-}=\pi,\nonumber
\end{align}
which corresponds to parallel field configuration. For the same sign tunneling amplitudes we obtain
\begin{equation}
\mc{A}_{-}=t
\begin{pmatrix}
1 & -a \\
1 & a
\end{pmatrix}=
R_{l}^{\dag}t
\begin{pmatrix}
\sqrt{2(1+a^2)} & 0 \\
0 & 0
\end{pmatrix}R_{d,+},
\end{equation}
\begin{equation*}
R_{d,+}=\frac{1}{\sqrt{1+a^2}}
\begin{pmatrix}
+1 & +a \\
-a & +1
\end{pmatrix},
\quad
\theta_{+}=-\arctan\left(\frac{2a}{1-a^2}\right).
\end{equation*}
Lastly, for tunneling amplitudes examined in \figurename~\ref{fig5} we get:
\begin{equation}
\mc{A}=t_{0}
\begin{pmatrix}
-\sqrt{0.3} & 1 \\
\sqrt{0.1} & \sqrt{0.4}
\end{pmatrix}\approx
R_{l}^{\dag}t_{0}
\begin{pmatrix}
1.23 & 0 \\
0 & 0.54
\end{pmatrix}R_{d},
\end{equation}
\begin{equation*}
R_{l}\approx\begin{pmatrix}
0.91 & 0.41 \\
-0.41 & 0.91
\end{pmatrix},\quad
R_{d}\approx\begin{pmatrix}
-0.30 & 0.95 \\
0.95 & 0.30
\end{pmatrix},
\quad
\theta\approx0.61.
\end{equation*}
In all above cases we have $\ve=-V_{g}$, $h=\Delta{E}/2$.

%\bibliography{refs_gediminas}

\begin{thebibliography}{80}
\bibitem [{Foo()}]{Footnote}%
  \BibitemOpen
  \href@noop {} {\bibinfo  {journal} {For example, after using typical
  parameters $U=5\Gamma$, $a=1$ and tunneling amplitudes in Eq.~(13) provides
  at a particle-hole symmetric point $V_{g}=0$ such Kondo temperature
  $k_{\mathrm{B}}T_{\mathrm{K}}=\sqrt{2U\Gamma}/\pi\exp\left[-\pi
  U/(8\Gamma)\right]\approx 0.14\Gamma$, which is substantially lower than the
  temperature used in our calculations, $k_{\mathrm{B}}\overline{T}=\Gamma$}\
  }\BibitemShut {NoStop}%
\end{thebibliography}

%

\end{document}